\documentclass[conference]{IEEEtran}
\IEEEoverridecommandlockouts
\usepackage{cite}
\usepackage{amsmath,amssymb,amsfonts}
\usepackage{algorithmic}
\usepackage{graphicx}
\usepackage{textcomp}
\usepackage{listings}
\usepackage{caption}
\usepackage{enumitem}
\usepackage{booktabs}
\usepackage{multirow}
\usepackage{float}
\usepackage[hidelinks]{hyperref}
\usepackage{titlesec}
\usepackage{threeparttable}  
\usepackage{tablefootnote}
\usepackage{indentfirst}

\titlespacing*{\section}
{0pt}{1.0\baselineskip}{0.5\baselineskip}

\titlespacing*{\subsection}
{0pt}{0.75\baselineskip}{0.25\baselineskip}

\titlespacing*{\subsubsection}
{0pt}{0.5\baselineskip}{0pt}
\usepackage{tabularx}
\usepackage[linesnumbered,ruled,vlined]{algorithm2e}
\usepackage{xcolor}
\def\BibTeX{{\rm B\kern-.05em{\sc i\kern-.025em b}\kern-.08em
    T\kern-.1667em\lower.7ex\hbox{E}\kern-.125emX}}
\begin{document}

\title{Data Race Satisfiability on Array Elements}

\author{\IEEEauthorblockN{Junhyung Shim}
\IEEEauthorblockA{\textit{Dept. of Computer Science} \\
\textit{Iowa State University}\\
Ames, Iowa, USA \\
jshim@iastate.edu}
\and
\IEEEauthorblockN{Quazi Ishtiaque Mahmud}
\IEEEauthorblockA{\textit{Dept. of Computer Science} \\
\textit{Iowa State University}\\
Ames, Iowa, USA \\
mahmud@iastate.edu}
\and
\IEEEauthorblockN{Ali Jannesari}
\IEEEauthorblockA{\textit{Dept. of Computer Science} \\
\textit{Iowa State University}\\
Ames, Iowa, USA \\
jannesar@iastate.edu}
}

\maketitle

\begin{abstract}
Detection of data races is one of the most important tasks for verifying the correctness of OpenMP parallel codes. Two main models of analysis tools have been proposed for detecting data races: dynamic analysis and static analysis. Dynamic analysis tools such as Intel Inspector, ThreadSanitizer, and Helgrind$^+$ can detect data races through the execution of the source code. However, source code execution can be quite time-consuming when analyzing computation-intensive programs. There are also static analysis tools such as LLOV, and OpenRace. These tools statically detect data races using algorithms that often do not require the execution of the source code. Although both detection techniques assist programmers in analyzing the correct behavior of OpenMP programs, they still produce false positives that often defeat the purpose of applying automatic analysis. Therefore, we present DRS-oNE (Data Race Satisfiability on aNy Element), a data race detector that detects data races on array elements by solving for race constraints with the Z3 SMT solver. DRS-oNE is available at \url{https://anonymous.4open.science/r/ompdart-DRS-extension-4152/README.md
}

\end{abstract}

\renewcommand{\IEEEkeywordsname}{Index Terms} 
\begin{IEEEkeywords}
\raggedright Parallel programming, OpenMP, Data race, Data race detection, Static Analysis
\end{IEEEkeywords}

\section{INTRODUCTION}
Recently, Kadosh et al. \cite{Kadosh_2023} reported that about 45$\%$ of HPCorpus dataset utilize OpenMP libraries. While only 0.16$\%$ - 0.54 $\%$ of all C/C++ for loops in the dataset were parallelized \cite{Kadosh_2023}, correct parallelization of such for loops is still necessary to guarantee consistent computation. One important factor of consistent computation under parallelization is ensuring that the parallel region is data race-free.

A data race is said to occur when two or more threads try to access the same memory location simultaneously, and at least one access is a write. It is relatively easier to detect data races on variables within a parallelized loop than that of array elements, as one just needs to find at least one read-write or write-write pair of a variable. However, detecting data races on array elements may not be so apparent once the indexing involves integer arithmetic operations. Listing \ref{lst:example1} shows an example of the mentioned problem. Once a thread works on $i=2$, and simultaneously, another thread works on $i=7$, then there is a data race on arr[14]. 

\begin{figure}[h]
\captionsetup{labelformat=default, labelsep=colon, name=Listing}
\begin{lstlisting}[language=C,basicstyle=\normalsize\ttfamily]
int arr[1000];
#pragma omp parallel for
for(int i = 0; i < 10; i++){
    arr[i%6 + 6*i] = arr[2*i];
}
\end{lstlisting}
\caption{Data race on an array element}
\label{lst:example1}
\end{figure}

Dynamic analysis tools \cite{intel_inspector,serebryany2009threadsanitizer, Helgrind+,helgrindplus-lib} are effective in detecting such data races, but,
dynamic detection may result in a long execution time or may fail to detect data races due to the manifestation of nondeterministic behavior of thread interleavings during the execution of the code. 

Static analysis tools avoid these issues by statically analyzing the textual representation of the source code. Recent static analysis data race detection tools such as LLOV and OpenRace have been reported to achieve precisions\footnote{precision = $\frac{TP}{TP + FP}$} of 0.91 \cite{Bora/taco/2020} and 0.89 \cite{openrace}, respectively. Since precision indicates the ratio of true positives out all all positive detections (true positives + false positives), there still exists a need to eliminate false positives for the mentioned tools. 

False positives occur in LLOV mainly due to relying on polyhedral model \cite{polly} to detect data races which require parallel loops to be affine. In the case of non-affine loops, LLOV relies on LLVM's Alias Analysis (AA) \cite{LLVM_AA}, but due to AA's conservative nature, it is possible for LLOV to create false positives.

OpenRace \cite{openrace} deploys modular, domain-specific analysis techniques to accurately detect data races. Currently, OpenRace consists of five stages: Preprocessing, Pointer Analysis, Function Summarization, Trace Building, and Analysis. However, the authors of OpenRace have stated that OpenRace ignores paths to avoid the path explosion problem. Due to this design choice, false positives occur when there is a data race inside an infeasible path. In this work, we Identify false positives reported by LLOV and OpenRace using a subset of DataRaceBench 1.4.0 \cite{dataracebench} along with a set of seven handwritten test cases to show that the race detection using constraint solving is effective in reducing false positive detections. 

We, therefore, present DRS-oNE, a static analysis tool that detects data races on array elements. DRS-oNE is control flow aware such that it can avoid false positives once control statements involve integer variables or macros that can be determined statically. DRS-oNE can effectively avoid false positives by checking the constraints required for the read-write or write-write operations to occur. These constraints are referred to as race constraints as described in \cite{pathG}. DRS-oNE extends race constraints by assuring that certain values of integer variables must hold if they can be determined statically. This allows for modeling constraints that depend on the values of integer variables declared outside the loop.

DRS-oNE is built on top of OMPDart \cite{ompdart} to utilize its recordings of array access information and the hybrid data structure, AST-CFG, that combines an Abstract Syntax Tree (AST) with a Control Flow Graph (CFG). Using OMPDart's AST-CFG and array access information, we encoded source code information into race constraints that can be solved by the Z3 SMT solver \cite{z3SMT}. By checking for constraints retrieved from CFG with the Z3 SMT solver, our detection tool can avoid infeasible paths that can be evaluated statically, which ultimately reduces false positive detections. A detailed explanation of DRS-oNE is described in Sections IV and V.

Currently, our detection tool only supports analysis for direct access to the array elements in a loop parallelized either with ``\#pragma omp for" or ``\#pragma omp simd". That is, the array was declared statically (i.e., int arr[N];), and accessing the memory region is only done by indexing the array identifier (i.e., arr[i];). Furthermore, it does not support data race caused by data race on a scalar variable. We, therefore, used a subset of Data Race Bench 1.4.0 \cite{dataracebench}, consisting of 55 files\footnote{files with more than one parallel construct, nested parallelism, GPU-related directives, data races on scalar variables, and indirect accesses of arrays were removed} to evaluate the performance of its effectiveness in detecting data races on array elements. Using the subset of Data Race Bench and seven handwritten test cases, we show that race detection using our implementation of the augmented race constraints can effectively reduce false positives.

We make the following contributions in this paper:
\begin{itemize}[leftmargin=2em]
    \item Extension of OMPDart to support data race detection of array elements within a parallelized loop
    \item Openly available implementation of augmented race constraints that originates from Pathg \cite{pathG} 
    \item Benchmark results to show the effectiveness of our augmented race constraints through testing with a subset of Data Race Bench 1.4.0 \cite{dataracebench} and seven additional test cases.
\end{itemize}

The remainder of the paper is structured as follows. Section II provides the related work. Section III provides motivating examples. Section IV provides the definition of race constraints and how we augmented the constraints. Section V describes the implementation details of DRS-oNE. Section VI describes the benchmark results and compares them with the latest data race detection tools. Section VII showcases usage examples where DRS-oNE might be useful. Section VIII elaborates on the limitations of DRS-oNE and future works that needs to be done to improve data race detection using race constraints. Then, the paper concludes with section IX.

\section{RELATED WORK}

\subsection{Race Detection using Polyhedral Model}
The polyhedral model is a powerful mathematical framework that enables the representation of finite affine loops as polyhedrons. It can be used to perform loop optimizations to exploit the principle of locality in caches or be used to analyze loop-carried dependencies for arrays. Polly \cite{polly} can perform this type of analysis and apply suitable loop transformations or apply OpenMP constructs. LLOV \cite{Bora/taco/2020} is a static data race detection tool that modifies Polly \cite{polly} to detect data races of affine loops with OpenMP constructs. Yet, due to the polyhedral model's inherent limitation, which requires the loop to be affine, the polyhedral model alone cannot detect data races for loops with complex non-affine structures. Therefore, LLOV relies on LLVM's Alias Analysis (LLVM AA) \cite{LLVM_AA} for data race detection of non-affine loops. However, due to the conservative nature of LLVM AA, LLOV can produce false positives. We found that LLOV is prone to creating false positive detections for complex loops, such as non-affine nested loops or loops with control statements. DRS-oNE avoids such false positive detections by constructing augmented race constraints, which do not require the target loop to be affine.

\subsection{Race Detection through multiple Analyses }
OpenRace \cite{openrace} is an open-source framework that consists of five stages of analysis. Preprocessing, Pointer Analysis, Function Summarization, Trace Building, and Analysis.
Through a series of analyses, OpenRace analyzes shared pointers between threads, collects abstract operations, and runs program traces. Using the information collected from these analyses, OpenRace then runs Shared Memory, Happens-Before, and Lockset analyses to detect data races. This whole program static analysis technique allows OpenRace to detect a wide range of data races. However, to avoid the path explosion problem in their whole program analysis, OpenRace ignores execution paths from control statements. This leads to the problem of incorrect analysis of infeasible paths, causing false positive data race detections. DRS-oNE can avoid such problem by constructing constraints to correctly identify infeasible paths.

\subsection{Race Detection using LLMs}
Chen et al. presented prompt engineering and fine-tuning techniques \cite{LeChenDataRaceLLM} for LLM models such as GPT-4 \cite{openai2024gpt4technicalreport}, StarChat- $\beta$ (based on StarCoder \cite{li2023starcodersourceyou}), and Llama-2 \cite{Llama2} to detect data races and showed the viability of using these LLMs for race detections. As mentioned in \cite{LeChenDataRaceLLM}, LLMs could generate detailed explanations about why a particular data race occurs and thus guide the programmers to fix the error. However, as research and survey works suggest \cite{Hallcination-Haung, Xu2024HallucinationII, Hallucination3}, LLMs are prone to Hallucination, where they create seemingly true but actually incorrect information. This suggests that traditional analysis tools are still required to ensure a safe analysis for race detections, as LLM alone could cause hallucination problems. 

\subsection{Race Detection through Dynamic Analysis}
Dynamic Analysis tools such as ThreadSanitizier \cite{serebryany2009threadsanitizer}, Helgrind$^+$ \cite{Helgrind+}, and Intel Inspector \cite{intel_inspector} work by instrumenting the source codes and executing them to find the errors at runtime. ThreadSanitizer uses the Happens-Before relations with the Lockset algorithm to track down data races during runtime. Helgrind$^+$ also uses Happens-Before relations with the Lockset algorithm along with support for the correct handling of condition variables. Intel Inspector employs a series of Happens-Before and Lockset analyses along with Dynamic Binary Instrumentations to detect data races. These techniques require source codes to be compilable into executables in order to run the analysis, and therefore, it lacks support for source codes that cannot be compiled on their own.

\subsection{Race Detection through Constraint Solving and Symbolic Execution}
Yu et al. proposed Pathg \cite{pathG}, which checks whether an OpenMP parallel program yields a consistent computation. Pathg first detects data race segments by solving for race constraints. Race constraint refers to a set of conditions required for data races to occur, mainly on array elements. If any constraint cannot be satisfied, then Pathg reports the parallel region as data race-free. If the constraints can be satisfied, then Pathg performs symbolic execution of the parallel region guided by the information from race constraints to search for inconsistencies. This guidance allows for pruning out unnecessary symbolic executions on accesses that do not logically create data races. However, without the openly available implementation, it is unclear whether their construction of constraints can correctly eliminate infeasible branches, especially for those resulting from the truth values of variables residing outside the parallel region. DRS-oNE extends the concept of race constraints to check the effectiveness of constraint-based data race detection.

\section{MOTIVATING EXAMPLES}
LLOV and OpenRace achieve outstanding precisions, 0.91 and 0.89, respectively. Yet, these works focus on the detection of the true positives, and therefore, they do not handle false positives with the same level of concern. However, these false positives create additional burdens for programmers by increasing the number of cases that programmers need to analyze. One case that results in false positives for both tools is the presence of control statements. If a parallel loop contains read and write operations inside a control structure, LLOV and OpenRace could create false positive reports, especially when the control structure creates an infeasible path such that when a program actually executes, that part of the code is not executed. Listing \ref{lst:FP1} and \ref{lst:FP2} depict such cases. In both cases, the boolean expressions are evaluated to be false, and therefore, during the execution of the programs, they do not cause data races.

\begin{figure}[h]
\captionsetup{labelformat=default, labelsep=colon, name=Listing}
\begin{lstlisting}[language=C,basicstyle=\normalsize\ttfamily]
    #define N 100
    
    int size = 100;
    int a = N;
    int b = N*N;
    int arr[size];

    #pragma omp parallel for
    for(int i = 0; i < 99; i++){
        if(a == b){
            arr[i] = arr[i+1] + i;
        }
    }

\end{lstlisting}
\caption{FP1 - False Positive Case-1 resulting from Infeasible Path}
\label{lst:FP1}
\end{figure}

\begin{figure}[h]
\captionsetup{labelformat=default, labelsep=colon, name=Listing}
\begin{lstlisting}[language=C,basicstyle=\normalsize\ttfamily]
    int size = 100;
    int arr[size];
    
    #pragma omp parallel for
    for(int i = 0; i < 99; i++){
        arr[i] = arr[i+1] + i;
    }

\end{lstlisting}
\caption{Simplified View of What OpenRace would Analyze for FP1}
\label{lst:openracesimplified}
\end{figure}

\begin{figure}[h]
\captionsetup{labelformat=default, labelsep=colon, name=Listing}
\begin{lstlisting}[language=C,basicstyle=\normalsize\ttfamily]
    int size = 100;
    int arr[size];

    #pragma omp parallel for
    for(int i = 0; i < 99; i++){
        if(i == 100){
            arr[i] = arr[i+1] + i;
        }
    }

\end{lstlisting}
\caption{FP2 - False Positive Case-2 resulting from Infeasible Path}
\label{lst:FP2}
\end{figure}

Listing \ref{lst:FP1} shows an example where LLOV and OpenRace both report the code as having a data race. However, Since the evaluation of  ``a == b" is false, given ``a = N" and ``b = N*N", the data race does not happen. Since the loop involves a control statement that requires a comparison of a non-linear expression (N*N) with a linear expression (N), LLOV will default to LLVM's Alias Analysis. As mentioned, due to the conservative nature of Alias Analysis, LLOV marks the loop as having a data race. 

OpenRace, on the other hand, marks Listing \ref{lst:FP1} as having a data race as it does not consider execution paths. Therefore, OpenRace would view Listing \ref{lst:FP1} as if it were Listing \ref{lst:openracesimplified}, where the control statement is removed. This, of course, results in false positive detection because it does not fully take into account the context where the read/write operations occur. Consequently, this mechanism of OpenRace causes OpenRace to report Listing \ref{lst:FP2} as having a data race as well, whereas LLOV can correctly avoid it as the loop is affine.

In contrast to LLOV and OpenRace, ThreadSanitizer can avoid these false positive cases as it executes the binary from the source code to check for data races. However, ThreadSanitizer may not be useful when the source code is difficult to run directly, i.e., it is not in the main function, whereas static analysis tools generally do not need to be compiled and run. Therefore, it is crucial to have a static analysis tool such that it can avoid false positive detections. To the best of our knowledge, DRS-oNE is the first static analysis tool that attempts to avoid false positive detections by solving for augmented race constraints.

\section{RACE CONSTRAINT}
Yu et al. \cite{pathG} proposed the concept of race constraint to identify regions that could produce a data race such that it can guide their symbolic execution model to check for inconsistencies. Race constraints in the original work consist of conditions that can be generalized into three main categories: (1) parallel conditions, (2) path conditions, and (3) conditions on the array indices. A parallel condition specifies the possible interval of the induction variable and the threads that can work on specific iteration values of the induction variable. The path condition is a conjunction of conditions that must be held for a read or a write operation to be executed, e.g., conditions of the surrounding if-else statements. Lastly, the condition on array indices is a condition of two different threads accessing the same memory address, and at least one access is a write. Using Listing \ref{lst:loopcntrl1}, we describe how the race constraints are formulated in the original work. Then, using Listing \ref{lst:loopcntrl2}, we show how our work models the augmented race constraints.  

\begin{figure}[h]
\captionsetup{labelformat=default, labelsep=colon, name=Listing}
\begin{lstlisting}[language=C,basicstyle=\normalsize\ttfamily]
int arr[1000];
#pragma omp parallel for
for(int i = 0; i < 10; i++){
    if(i < 5){
        arr[i%6 + 6*i] = arr[2*i];
    }
}

\end{lstlisting}
\caption{Loop with Control Statement}
\label{lst:loopcntrl1}
\end{figure}

In the original work, the race constraint of the array accesses in Listing \ref{lst:loopcntrl1}  would be formulated as:\\

$1)\indent divisible(i_1-0,m) \land divisible(i_2-1,m)$\\
\indent $2) \indent \land \enspace(0 \leq i_1 < 10)\land (0 \leq i_2 < 10)$\\
\indent $3) \indent \land \enspace (i_1 < 5) \land (i_2 < 5)$\\ \indent$4) \indent \land \enspace (i_1\%6+6*i_1) = (2*i_2)$\\

The parallel condition is given in 1) and 2), path condition in 3), and conditions on array indices in 4). Here, $m$ is the number of threads that execute the parallel loop. In this example, we consider two logical threads working on the loop, hence $m=2$. Assuming a static scheduling scheme with chunk size set to one, $divisible(i_1-0,m)$ ensures that $i_1$ is worked on by thread 0. Similarly, $divisible(i_2-1,m)$ ensures $i_2$ is worked on by thread 1. In the case of dynamic scheduling, pathG's race constraints are set up such that the number of threads is equal to the number of iterations, modeling each thread working only on one iteration. 

The formula above described in plain English would be: ``Given two threads working on the iteration values ($i_1,i_2$) that they need to execute, is it possible for those two threads to access the same memory address?" If the satisfiability solver can find a pair of $i_1,i_2$ such that the given constraints can be satisfied, then there is a data race.

Our augmented version of the race constraint consists of four conditions: (1) parallel condition, (2) path condition, (3) conditions on array indices, and (4) conditions on integer variables. Parallel condition in our work always assumes the dynamic scheduling scheme so that the analysis does not have to depend on the number of threads, which often varies from system to system. Conditions (2) and (3) follow the original race constraint. Conditions on integer variables are only applicable to those that can be determined statically. We show how our constraint is constructed from the example in Listing \ref{lst:loopcntrl2}. In the remainder of this paper, we will refer to race constraints as our augmented race constraints. 

Our race constraint from Listing \ref{lst:loopcntrl2} is constructed as:\\

\indent\indent $1) \indent  i_1 \neq i_2$\\
\indent\indent $2) \indent \land \enspace(0 \leq i_1 < 10)\land (0 \leq i_2 < 10)$\\
\indent\indent $3) \indent \land \enspace (i_1 < 5) \land (i_2 < 5)$\\ 
\indent\indent $4) \indent \land \enspace (i_1\%a+a*i_1) = (2*i_2)$\\
\indent\indent $5) \indent \land \enspace a = 6$\\

\begin{figure}[h]
\captionsetup{labelformat=default, labelsep=colon, name=Listing}
\begin{lstlisting}[language=C,basicstyle=\normalsize\ttfamily]
int arr[1000];
int a = 6;
#pragma omp parallel for
for(int i = 0; i < 10; i++){
    if(i < 5){
        arr[i%a + a*i] = arr[2*i];
    }
}
\end{lstlisting}
\caption{Loop with Control Statement and a Variable}
\label{lst:loopcntrl2}
\end{figure}

\noindent The parallel condition consists of 1) and 2), the path condition consists of 3), the condition on array indices corresponds to 4), and the condition on integer variable corresponds to 5). By incorporating the condition on the integer variables, we can model the race constraints more accurately. This contributes to the elimination of false positive data race detections caused by infeasible paths. However, such a condition is only added if the variable's value can be determined statically, i.e., the most recent update to the variable was an assignment of integer literal.

\section{IMPLEMENTATION}
\subsection{Usage}
The current version of DRS-oNE only supports the parallelization of one loop and requires the user to specify the target loop with ``\#pragma drs." The analysis performed by DRS-oNE can be generalized to performing a data race analysis on``\#pragma omp parallel for" or ``\#pragma omp simd". It is currently meant to check if there exists a data race before applying the parallel constructs. We leave the work on enabling multiple loop analyses with different parallel constructs for future work. 
\setcounter{figure}{0}
\begin{figure*}[t]
    \centering
    \includegraphics[width=\textwidth]{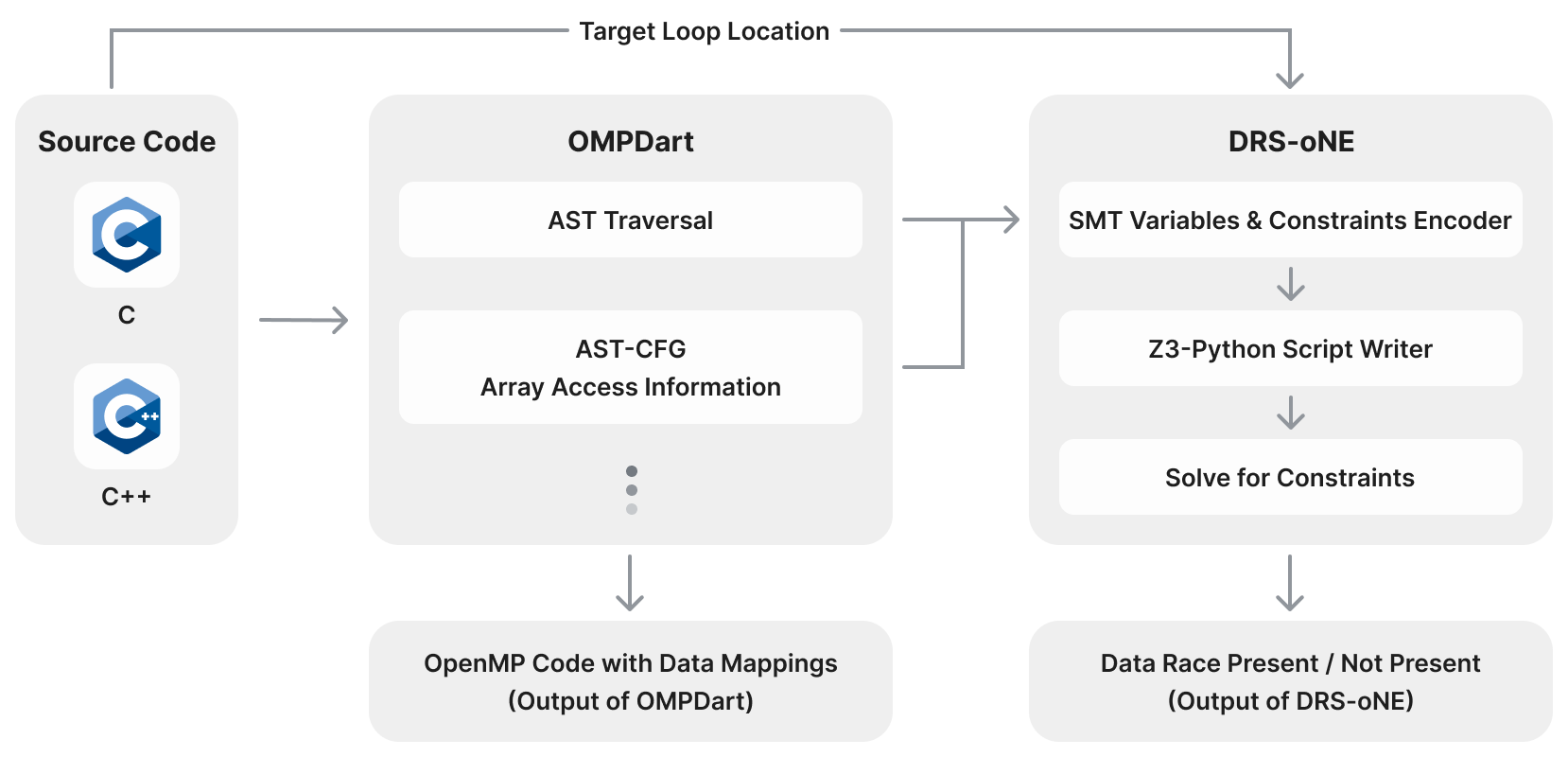}
    \caption{Workflow diagram of DRS-oNE}
    \label{fig:wide_figure}
\end{figure*}
\subsection{Workflow of DRS}
DRS is built on top of OMPDart \cite{ompdart} to utilize its hybrid data structure and its array access information. Figure 1 shows the diagram of DRS-oNE's workflow. DRS-oNE consists of four main stages. 
The first stage (topmost horizontal arrow) locates the target loop location. Then, the second stage encodes the source code information into SMT variables and constraints. The encoding of the variables, however, is partially done during the traversal of the AST, which is part of the OMPDart's procedures. After encoding variables and constructing constraints, a procedure writes a Python script that will be used to solve for the given variables and constraints. After a script is written and executed, DRS-oNE reports whether a data race is present in the loop. However, it does not report which type of data race it is, e.g., write after write, read after write, and write after read. The following subsections will describe each procedure of DRS.

\subsection{Locating the Target Loop}
DRS-oNE locates the target loop subject to analysis with the help of the Clang frontend's pragma handler. The pragma handler searches for ``\#pragma drs" and marks the source code line number of the pragma. Then, DRS-oNE searches if there is a for-loop that is located right underneath the pragma. If a for-loop is found, then DRS-oNE marks that loop as the target loop. As mentioned, the current implementation of the DRS-oNE does not support multiple target loops.
\subsection{SMT Variables \& Constraints Encoder}
The encoding of SMT variables and constraints in DRS-oNE mainly consists of four sub-procedures: (1) encoding of variables declared before the target loop, (2) encoding of the loop induction variables and their constraints, (3) encoding of the array indices and their constraints, and (4) construction of race constraints using the information from (1), (2), and (3).  

Encoding of variables declared before the target loop occurs during the traversal of the AST in OMPDart. As described in Algorithm 1, when a variable is declared, we store it in the map identified as $allVars$ to record that a variable has been declared. In the map, each variable can have its value set to an integer literal if the latest assignment to that variable was an integer literal. If the variable gets updated through some operation that isn't an assignment of an integer literal, then we mark that variable as ``no value" to signal that the value cannot be determined. This determination of integer variables' values ultimately allows DRS-oNE to perform an accurate analysis of feasible paths, which results in the elimination of false positives reported by other static analysis tools. However, DRS-oNE is not aware of variable updates through a pointer. Thus, it is possible that a variable in the map could have a false literal value if that variable gets updated through a pointer.

\SetAlgoNoEnd

\begin{algorithm}[h]
\SetAlgoLined
\KwIn{AST \(A\) of the source code}
\KwOut{Map \(allVars\) that contains the variables' information}

\SetKwFunction{recrdVar}{\textbf{record\(\_\)variables}}
\SetKwProg{Fn}{Function}{:}{}
\Fn{\recrdVar{\(A\)}}{
    \(allVars \gets \{\}\)\;
    \For{every node \(n\) in \(A\)}{
        \If{\(n\) is a variable declaration}{
            \(name \gets n.get\_variable\_name()\)\;
            \If{\(n\) gets assigned with integer literal \(I\)}{
                \(allVars[name] \gets I\)\;
            
            }\Else{
                \(allVars[name] \gets \text{no value}\)\;
            }

        }\ElseIf{\(n\) is an update of a variable}{
             \(name \gets n.get\_variable\_name()\)\;
            \If{\(n\) updates the variable with assignment of an integer literal \(I\)}{
                \(allVars[name] \gets I\)\;
            
            }\Else{
                \(allVars[name] \gets \text{no value}\)\;
            }
        }
    }
    \Return \(allVars\)\;
}
\textbf{End Function}
\caption{Encoding of Variables Declared before the Target Loop}
\end{algorithm}

The encoding of the loop induction variables and their conditions occurs after OMPDart has finished constructing its AST-CFG. As described in Algorithm 2., If we encounter a for-loop node in the AST-CFG, we extract the loop induction variable and store it into the map $indexVar$. It is crucial to know which variables are loop induction variables, as they are used to construct the parallel condition. The $get\_conditions()$ sub-procedure returns the loop bounds of the current loop and that of the parent loops that encapsulate the current loop. Furthermore, if the current loop is the target loop subject to parallelization, then the condition enforces that any two accesses using the induction variable from the target loop only happen when the values of the induction variable are different, as we assume that each thread only works on one unique iteration (this is equivalent to setting up the parallel region id in \cite{pathG}).

The encoding of the array indices and their constraints are described in Algorithm 2. and Algorithm 3. When DRS-oNE encounters an array access node in the AST-CFG, it records the array access indices and specifies the access type, which is either read or write. An assignment operation to an array will result in marking that array access as a write, and other operations will result in marking the access as a read. Compound assignment operation on an array, such as ``+=", will map the array access as a write. 

The ``$indices.encode()$" sub-procedure in Algorithm 3. encodes the indices of an array that is being accessed. For example, it will encode $arr[i+1][j*2][k]$ in the following manner: (1) Index of each dimension is created, i.e., $index\_1,$ $ index\_2,$ $ index\_3$. (2) each $index\_n$ gets the expression that is used to index the $n$-th dimension, i.e., $index\_1 == (i+1)$, $index\_2 == (j*2)$, and $index\_3 == k$.  (3) The target loop's induction variable is substituted with a copy of that induction variable, e.g., if the target loop's induction variable is $i$, then $index\_1 == (i+1)$ becomes $index\_1 == (i' + 1)$. This is done to enforce the parallel condition, as using the same variable for multiple accesses will always result in an unsatisfiable parallel condition ($i \neq i$ is never satisfiable). (4) lastly, to differentiate each access, we label each $index\_n$ and the substituted induction variables with a unique id. Therefore, the encoding of indices becomes: $index\_1\_0 == (i'\_0)$, $index\_2\_0 == (j+1)$, and $index\_3\_0 == (k)$.

The ``$get\_access\_conditions()$" sub-process retrieves and encodes the conditions required for access to the current indices to occur. In our implementation, we only support the condition of if statements. For if and else-if blocks, the logical conditions that lead to that access are directly added to the constraints. If the access occurs in an else block, then the conjunction of negation of the logical conditions of the control structure is added. After adding the conditions, they are encoded following the encoding schemes of (3) and (4) of ``$indices.encode()$".

\begin{algorithm}[h]
\SetAlgoLined
\KwIn{AST-CFG \(AC\) retrieved from OMPDart, \(allVars\) from Algorithm 1.}

\SetKwFunction{recrdVar}{\textbf{record\(\_\)idx\(\_\)vars\(\_\)and\(\_\)conds}}
\SetKwProg{Fn}{Function}{:}{}
\Fn{\recrdVar{\(AC, allVars\)}}{
    \(inductionVars \gets \{\}\)\;
    \(readMap \gets \{\}\)\;
    \(writeMap \gets \{\}\)\;
    \For{every node \(n\) in \(AC\)}{
        \If{\(n\) is a loop}{
            \(induction\_var \gets n.get\_induction\_variable()\)\;\
            
            \(condition \gets n.get\_conditions()\)\;\
            
            \(inductionVars[induction\_var] \gets condition\)\;
        }\ElseIf{\(n\) is an array access}{
            \(indices \gets n.get\_array\_indices()\)\;
            \(access\_type \gets n.get\_access\_type()\)\;
            \(record\_indices(
            indices, access\_type, \newline write\_list, read\_list)\)\;
        }
    }
    
}
\textbf{End Function}
\caption{Encoding of Array Indices}
\end{algorithm}

\begin{algorithm}[h]
\SetAlgoLined
\KwIn{\(indices\), \(access\_type)\), \(write\_list\), \(read\_list\) from Algorithm 1. and 2.}

\SetKwFunction{recrdVar}{\textbf{record\(\_\)indices}}
\SetKwProg{Fn}{Function}{:}{}
\Fn{\recrdVar{\(indices, access\_type,\newline write\_list, read\_list\)}}{
    \( indices \gets indices.encode()\) \;\
    
    \(access\_conditions \gets indices.get\_access\_conditions()\)\;\

    \(encodings \gets (indices + access\_conditions )\)\;\
    
    \If{\text{access\_type is write}}{
        \(write\_list.append(encodings)\)
    }\ElseIf{\text{access\_type is read}}{
        \(read\_list.append(encodings)\)
    }
}
\textbf{End Function}
\caption{Encoding of Induction Variables and Array Accesses}
\end{algorithm}

\subsection{Z3-Python Script Writer}

The script writer stage is responsible for writing the Python script that will solve the constraints using the Python Z3-SMT library. The script consists of constraints for detecting data races from RAW (Read after Write) and WAW (Write after Write) dependencies, which are constructed by pairing $write\_list$ with $read\_list$ and with itself. Write after Read dependencies are not considered as it is equivalent to RAW dependencies in the loop.

\section{BENCHMARKS}
Since DRS-oNE only supports data race detection on array elements under SIMT (Single instruction, multiple threads) settings, we performed the benchmark on the subset of DataRaceBench 1.4.0. \cite{dataracebench}. The subset excludes SIMD, GPU, and multiple parallel constructs (since DRS-oNE currently only supports one parallel construct) from the original dataset. We compare the results against LLOV \cite{Bora/taco/2020}, OpenRace \cite{openrace}, and ThreadSanitizer \cite{serebryany2009threadsanitizer}. Additionally, we compare DRS-oNE against the mentioned tools on seven different handwritten loops to observe the effectiveness of DRS-oNE in eliminating false positives triggered by infeasible paths.

\begin{table}[ht]
\centering
\normalsize
\caption{Data Race Detection Tools Comparison}
\label{tab:race_detection}
\begin{threeparttable}
\begin{tabularx}{\linewidth}{|X|c|c|c|c|c|}
\hline
\textbf{Tools} & \multicolumn{2}{c|}{\textbf{Race: yes}} & \multicolumn{2}{c|}{\textbf{Race: no}} & \textbf{Coverage/55} \\
\cline{2-5}
               & \textbf{TP} & \textbf{FN} & \textbf{TN} & \textbf{FP} & \\
\hline
LLOV           & 29          & 1           & 20          & 2           & 52 \\
OpenRace       & 31          & 0           & 24          & 0           & 55 \\
ThreadSanitizer$^*$ & 19        & 13          & 17          & 7           & 55 \\
\textbf{DRS-oNE}   & 26          & 4          & 24          & 0           & 54 \\
\hline
\end{tabularx}
\begin{tablenotes}
        \item $^*$Averaged over 5 runs. Due to ThreadSanitizer's dynamic analysis technique, the result may differ each time the benchmark is run.
    \end{tablenotes}
    \end{threeparttable}
\end{table}

\begin{table}[ht]
\centering
\normalsize
\caption{Performance Metrics}
\label{tab:perfmetrics}
\begin{tabularx}{\linewidth}{|X|c|c|c|c|}
\hline
\textbf{Tools}   & \textbf{Precision} & \textbf{Recall} & \textbf{Accuracy} & \textbf{F1} \\
\hline
LLOV            & 0.935              & 0.967           & 0.942             & 0.951       \\
OpenRace        & 1.000              & 1.000           & 1.000             & 1.000       \\
ThreadSanitizer & 0.731              & 0.613           & 0.655             & 0.667       \\
\textbf{DRS-oNE}    & 1.000              & 0.867           & 0.926             & 0.929       \\
\hline
\end{tabularx}
\end{table}

Table \ref{tab:race_detection} displays the benchmark results on the subset of DataRaceBench 1.4.0. LLOV and OpenRace statically analyze the source codes, whereas ThreadSanitizer dynamically analyzes them. Table \ref{tab:perfmetrics} displays the common statistics used to evaluate data race detection tools. 

From Table II, it can be observed that DRS-oNE has low accuracy compared to other static analysis tools (LLOV and OpenRace). This mainly results from two reasons: (1) DRS-oNE's inability to detect data race caused by indirect accesses, e.g., pointers and function calls, and (2) its inability to detect data races on variables. Despite such low detection rate, DRS-oNE achieves comparable precision to that of OpenRace and ThreadSanitizer, indicating that it can effectively avoid false positive detections. 

From Table \ref{tab:race_detection}, it can be observed that LLOV produced three false positives. Evaluating the three source codes revealed that only one of them was correctly avoided by DRS-oNE; The others were not reported due to DRS-oNE's inability to support function calls and runtime-dependent inputs. We, therefore, show how DRS-oNE was effective in eliminating that false positive case. Listing  \ref{lst:drb054} is an excerpt of the source code from DRB054-inneronly2-orig-no.c, which LLOV reported as having a data race. This file, however, does not have a data race because the inner for-loop has an implicit barrier at the end of the loop according to the specification of OpenMP 6.0 \cite{openmp60}. This means that the parallel execution of the inner loop must finish before the outer loop proceeds to the next iteration. Hence, there is no data race. Yet, LLOV marks b[i][j]=b[i-1][j-1] as having a data race because its data race detection algorithm computes the Reduced Dependence Graph\footnote{which is constructed by utilizing Polly} (RDG) only on regions marked as parallel \cite{Bora/taco/2020}. DRS-oNE can correctly identify that there is no data race for this case because the constraints are aware that the outer loop is non-parallel, and hence, the iteration values of the outer loop's induction variable cannot be different between different threads. This means that the indices of the first dimension of b from the left and right-hand sides are always different during the execution of the parallel loop, which ensures that there is no data race.

\setcounter{figure}{6}
\begin{figure}[h]
\captionsetup{labelformat=default, labelsep=colon, name=Listing}
\begin{lstlisting}[language=C,basicstyle=\normalsize\ttfamily]
for (i=1;i<n;i++)
  #pragma omp parallel for
  for (j=1;j<m;j++)
    b[i][j]=b[i-1][j-1];
\end{lstlisting}
\caption{DRB054-inneronly2-orig-no.c}
\label{lst:drb054}
\end{figure}

OpenRace achieves phenomenal performance metrics in this subset benchmark. However, it lacks the ability to distinguish execution paths as it does not account for conditional branching \cite{openrace}. This may result in false positives if the race happens conditionally, e.g., in the infeasible path. We, therefore, compare DRS-oNE with OpenRace along with other tools to evaluate their effectiveness in avoiding false positive cases induced by infeasible paths. Table III shows the comparison results. In the table, D denotes that a data race was reported, and ND denotes that no data race was reported. We provide the falsely detected source codes in Listing \ref{lst:fp3}, Listing \ref{lst:fp6}, and Listing \ref{lst:fp7} (FP1 and FP2 can be found in Listing \ref{lst:FP1} and Listing \ref{lst:FP2}, respectively, under section III). The rest of the test files can be found in the source code in the ``/FalsePostiveBench" sub-directory.

\begin{table}[h]
\normalsize
\centering 
\captionsetup{justification=centering, labelsep=colon} 
\caption{Comparison of Tools on False Positive Detections, (TSN = ThreadSanitizer)}
\label{tab:fpDetection}
\begin{tabularx}{\linewidth}{|X|c|c|c|c|}
\hline
\textbf{Files/Tools} & \textbf{OpenRace} & \textbf{LLOV} & \textbf{DRS-oNE} & \textbf{TSN} \\
\hline
\textbf{FP1} & D & D & D & ND \\
\hline
\textbf{FP2} & D & ND & ND & ND \\
\hline
\textbf{FP3} & D & ND & ND & ND \\
\hline
\textbf{FP4} & ND & ND & ND & ND \\
\hline
\textbf{FP5} & ND & ND & ND & ND \\
\hline
\textbf{FP6} & D & D & ND & ND \\
\hline
\textbf{FP7} & D & D & ND & ND \\
\hline
\hline
\textbf{Total} & 5/7 & 3/7 & 1/7 & 0/7 \\
\hline
\end{tabularx}
\end{table}

\begin{figure}[h]
\captionsetup{labelformat=default, labelsep=colon, name=Listing}
\begin{lstlisting}[language=C,basicstyle=\normalsize\ttfamily]
    int size = 100;
    int arr[size];
    
    #pragma omp parallel for
    for(int i = 0; i < 10; i++){
        arr[i] = arr[i%10] + i;
    }
\end{lstlisting}
\caption{False Positive Case 3 (FP3)}
\label{lst:fp3}
\end{figure}

\begin{figure}[h]
\captionsetup{labelformat=default, labelsep=colon, name=Listing}
\begin{lstlisting}[language=C,basicstyle=\normalsize\ttfamily]
    #define N 100
    
    int size = 100;
    int a = 0;
    int b = N;
    int arr[size];

    #pragma omp parallel for
    for(int i = 0; i < N; i++){
        if(a == 0 && b != N){
            arr[i] = arr[i] + 1;
            arr[i] = arr[i+1] + 1;
        }
    }
\end{lstlisting}
\caption{False Positive Case 6 (FP6)}
\label{lst:fp6}
\end{figure}

\begin{figure}[h]
\captionsetup{labelformat=default, labelsep=colon, name=Listing}
\begin{lstlisting}[language=C,basicstyle=\normalsize\ttfamily]
    #define N 100
    
    int size = 100;
    int a = 0;
    int b = N;
    int arr[size];

    #pragma omp parallel for
    for(int i = 0; i < N; i++){
        arr[i] = arr[i] + 1;
        arr[i] = arr[i+1] + 1;
        
    }
\end{lstlisting}
\caption{OpenRace's view on False Positive Case 6 (FP6), Simplified}
\label{lst:fp6-omprace-view}
\end{figure}

\begin{figure}[h]
\captionsetup{labelformat=default, labelsep=colon, name=Listing}
\begin{lstlisting}[language=C,basicstyle=\normalsize\ttfamily]
    int arr[1000];
    
    #pragma omp parallel for
    for(int i = 0; i < 10; i++){
        if(i*i*i >= 1000){
            arr[i%6 + 6*i] = arr[2*i];
        }
    }
\end{lstlisting}
\caption{False Positive Case 7 (FP7)}
\label{lst:fp7}
\end{figure}

\begin{figure}[h]
\captionsetup{labelformat=default, labelsep=colon, name=Listing}
\begin{lstlisting}[language=C,basicstyle=\normalsize\ttfamily]
    int arr[1000];
    
    #pragma omp parallel for
    for(int i = 0; i < 10; i++){
        arr[i%6 + 6*i] = arr[2*i];
    }
\end{lstlisting}
\caption{OpenRace's View on False Positive Case 7 (FP7), Simplified}
\label{lst:fp7-openrace-view}
\end{figure}

We can observe that DRS-oNE effectively avoids false positives that other static analysis tools cannot avoid. However, it can be observed that DRS-oNE failed to avoid FP1. As seen from Listing \ref{lst:FP2}, the control statement involves the evaluation of ``a == b". Although ``a" can be determined statically, the value of ``b" is considered dynamic by DRS-oNE, and, therefore, the constraint for b is not added (refer to Algorithm 1.) 

OpenRace reports Listing \ref{lst:fp3} as having a data race even though there aren't any control structures in the loop body. This could be due to a number of reasons, but we suspect that it is due to the conservative nature of shared-memory analysis computed from pointer analysis. Additionally, we can observe that OpenRace and LLOV report Listing \ref{lst:fp6} and Listing \ref{lst:fp7} as having data races. In both cases, as mentioned earlier, OpenRace reports them as having data races because it does not consider the path of execution, and thus, its analysis is equivalent to performing an analysis on the codes seen from Listing \ref{lst:fp6-omprace-view} and Listing \ref{lst:fp7-openrace-view} For LLOV, both cases are non-affine and so it defaults to LLVM Alias Analysis, which results in creating false positive reports. DRS-oNE, on the other hand, can effectively avoid false positives seen from Listing \ref{lst:fp6} and Listing \ref{lst:fp7} by correctly constructing constraints and solving them.

\section{USE CASES}
DRS-oNE's analysis can be generalized into detecting data races on arrays in a loop with ``\#pragma omp parallel for" or ``\#pragma omp simd", as it essentially performs loop-carried dependence on arrays using the Z3 SMT solver. This is particularly helpful when a programmer wants to parallelize a loop that performs operations on arrays. Consider Listing \ref{lst:histogram-usage}, where the user decides to parallelize the loop with ``\#pragma omp parallel for". The objective of this loop is to populate some parts of the histogram based on some arithmetic rules. Now, to ensure the correct parallelization, the programmer needs to figure out whether there exists a data race between two arbitrary array write statements within the loop, which is relatively hard to prove by hand in this case, especially with additional constraints imposed by the control statements. Thus, using DRS-oNE, the programmer can check for data races on arrays without going through the hassle of proving them by hand.   

\begin{figure}[h]
\captionsetup{labelformat=default, labelsep=colon, name=Listing}
\begin{lstlisting}[language=C,basicstyle=\normalsize\ttfamily]

    int histogramSize = 200000000;
    int histogram[histogramSize];

    for (int j = 0; j < histogramSize; j++) {
        histogram[j] = 0;
    }
    #pragma omp parallel for
    #pragma drs //needed for analysis
    for (int i = 0; i < 100; i++) {
        if((i*i*i)%2 == 0){
            histogram[(i*i*i)] = i;
        }else if(i % 5 == 0){
            histogram[((i+1)*
                       (i+1)*
                       (i+1)*
                       (i+1))] = 30*i;
        }else if(i % 13 == 0){
            histogram[((i-1)*
                       (i-1)*
                       (i-1)*
                       (i-1))] = 40*i;
        }else if (i > 1){
            histogram[((i-1)*
                       (i-1)*
                       (i+2)*
                       (i+3))] = 50*i; 
        }
    }

\end{lstlisting}
\caption{Example of where DRS-oNE's analysis would be useful}
\label{lst:histogram-usage}
\end{figure}

Another use case of DRS-oNE is to check whether other tools' output is indeed a true negative or not, as DRS-oNE is resilient to false positive detections. Furthermore, the avoidance of false positives may come in handy for verifying codes that were parallelized automatically. Consider a prompt in Listing \ref{lst:gpt-prompt} that was given to GPT-4o. For that prompt, GPT-4o responded with the answer seen from Listing \ref{lst:gpt-answer}.

\begin{figure}[t]
\captionsetup{labelformat=default, labelsep=colon, name=Listing}
\begin{lstlisting}[language=C,basicstyle=\normalsize\ttfamily]
    Q: Can you correctly Parallelize the following loop?
    
    #define N 100
    int size = 100;
    int a = N;
    int b = N*N;
    int arr[size];
    
    for(int i = 0; i < 10; i++){
        arr[i] = arr[i%10] + i;
    }
\end{lstlisting}
\caption{Prompt \#1 given to GPT-4o}
\label{lst:gpt-prompt}
\end{figure}

\begin{figure}[h]
\captionsetup{labelformat=default, labelsep=colon, name=Listing}
\begin{lstlisting}[language=C,basicstyle=\normalsize\ttfamily]
    A: 
    ...
    No, the loop cannot be directly 
    parallelized as it stands because
    of the dependency between 
    iterations. Each iteration depends 
    on the arr[i % 10] value, which 
    may be updated by another 
    iteration, leading to a race 
    condition.
    ...
\end{lstlisting}
\centering
\caption{Answer From GPT-4o Corresponding to the prompt in Listing \ref{lst:gpt-prompt}}
\label{lst:gpt-answer}
\end{figure}

 Based on the answer, GPT-4o thinks that there is a dependence between arr[i\%10], but because the loop bound is [0,9], arr[i\%10] is equivalent to arr[i] during the entire execution of the loop, which means that there is no dependence that causes a data race. Now, had the user tried to automatically parallelize the code in Listing \ref{lst:gpt-prompt} using LLM-based auto parallelization techniques (such as AutoParLLM \cite{Mahmud2023AUTOPARLLMGA}), and the user tried to use OpenRace for the verification of the output to see if it agrees with the LLM's output, the user would have missed the parallel opportunity as the LLM and OpenRace would both have invalidated it due to their conservative dependence analyses. In this case, DRS-oNE can be used to alert the user that the verification tool resulted in a false positive, and needs to be investigated.
 
 Another similar use case would be to guide LLMs into generating correct parallelization. It is possible that the LLMs miss parallelization opportunities due to the incorrect evaluation of the paths. The prompt in Listing \ref{lst:gpt-prompt2} asks GPT-4o to parallelize the loop. This loop does not have a data race because the read and write operations in the loop do not occur as ``b $!=$ N" evaluates to false. Therefore, it can be parallelized (although its computation is meaningless). However, The response from GPT-4o seen from Listing \ref{lst:gpt-ans2} states that the loop cannot be parallelized due to the dependencies within the loop. This suggests that the current 4o model has limited awareness of control flows. Therefore, DRS-oNE's constraints could be used as a starting point for guiding LLMs to correct their mistakes in parallelization.  
 \begin{figure}[h]
\captionsetup{labelformat=default, labelsep=colon, name=Listing}
\begin{lstlisting}[language=C,basicstyle=\normalsize\ttfamily]
    Q: Can you parallelize the following loop?
    
    #define N 100
    int size = 100;
    int a = 0;
    int b = N;
    int arr[size];

    for(int i = 0; i < N; i++){
        if(a == 0 && b != N){
            arr[i] = arr[i] + 1;
            arr[i] = arr[i+1] + 1;
        }
    }
\end{lstlisting}
\centering
\caption{Prompt \#2 given to GPT-4o}
\label{lst:gpt-prompt2}
\end{figure}

 \begin{figure}[h]
\captionsetup{labelformat=default, labelsep=colon, name=Listing}
\begin{lstlisting}[language=C,basicstyle=\normalsize\ttfamily]
    A:
    ...
    The given loop cannot be directly 
    parallelized because of 
    dependencies within the loop.
    ...
\end{lstlisting}
\centering
\caption{answer From GPT-4o Corresponding to the prompt
in Listing \ref{lst:gpt-prompt2}}
\label{lst:gpt-ans2}
\end{figure}

Lastly, DRS-oNE could also serve as a verification tool for those tools that automatically parallelize regions through the discovery of parallel opportunities \cite{Discopop, Pattern2, pattern3},  and automated OpenMP construct generation \cite{autmaticConstSelect} by verifying the existence of data races. Ideally, using a collection of various analysis tools that includes DRS-oNE would be useful as it can complement tools with high false positive detection rates, whereas other analysis tools can compensate for the current version of DRS-oNE's inability to detect various true positives.

\section{DISCUSSION}
DRS-oNE leverages the concept of race constraints introduced in \cite{pathG} and extends it by adding additional constraints. This paper shows that data race constraints are effective in detecting data races on array elements while avoiding false positive detections. However, unlike other static analysis tools, DRS-oNE lacks support for various OpenMP parallel constructs because it only logically supports ``\#pragma omp parallel for" and requires the user to manually specify the target loop. In our future work for DRS, we plan to remove these issues by properly handling parallel constructs. Furthermore, the static analysis at the Abstract Syntax Tree (AST) level hinders the development and support of various constraints, as it requires edge-case handling to properly encode source code text information into SMT variables and constraints. Therefore, in out future work, we plan to construct race constraints using LLVM-IR instead of using AST.

Additionally, the use of AST makes the process of modeling function calls and conditional branching difficult. Currently, DRS-oNE does not support function calls, as it is difficult to analyze function behaviors solely at the AST level. As for conditional branches, DRS-oNE only supports if-else statements due to the complexity of handling edge cases of constructing constraints using AST nodes. These problems, once again, demand the switch from AST to LLVM-IR, as it allows for the ease of scalability and reduces development time for the detection tool.

Another limitation of DRS-oNE is that it cannot detect data races on scalar variables and races caused by pointer access. Data races on scalar variables and pointers can propagate the race to array elements. In such cases, DRS-oNE would fail to detect the propagated data races. Therefore, in our future work, we plan to add support for constraints on scalar variables and pointer accesses as well.

Lastly, DRS-oNE constructs constraints using information that can be determined statically, e.g., macros and integer variables that were assigned with literal values. Although such information allows for the construction of more accurate constraints than those without it, it does not accurately reflect the behavior of programs used in practice, as many variables in a program typically depend on runtime computation. This demands the need to incorporate runtime constraints. 
One possible solution to this problem would be to incorporate dynamic invariant detection tools such as Daikon \cite{Daikon} to extract the constraints. Another solution is the use of AI Agents to predict the runtime behavior. Recently, Mahmud et al. proposed AutoParLLM \cite{Mahmud2023AUTOPARLLMGA}, which uses a GNN on a flow-aware graph representation of a source code (PerfoGraph \cite{perfograph}) to construct appropriate prompts for LLMs to parallelize the sequential code. We anticipate that using a GNN to extract the flow-aware information to construct dynamic constraints would be faster than dynamic invariant detection tools.

\section{CONCLUSION}
In this paper, we addressed the problem of static analysis tools reporting false positives due to their conservative analysis techniques. We proposed an augmented race constraint that incorporates augmented information and showed its viability of detecting data races and its effectiveness in avoiding false positive detections by running benchmarks on a subset of DataRaceBench 1.4.0 and seven handwritten test cases. We then showcased that DRS-oNE could prove to be useful when it is used to verify data race-free auto-parallelized codes with other analysis tools. These findings and discussions highlight the effectiveness of our augmented race constraints for accurate race detections that produce fewer false positives.

\bibliographystyle{IEEEtran}

\begin{thebibliography}{10}
\providecommand{\url}[1]{#1}
\csname url@samestyle\endcsname
\providecommand{\newblock}{\relax}
\providecommand{\bibinfo}[2]{#2}
\providecommand{\BIBentrySTDinterwordspacing}{\spaceskip=0pt\relax}
\providecommand{\BIBentryALTinterwordstretchfactor}{4}
\providecommand{\BIBentryALTinterwordspacing}{\spaceskip=\fontdimen2\font plus
\BIBentryALTinterwordstretchfactor\fontdimen3\font minus \fontdimen4\font\relax}
\providecommand{\BIBforeignlanguage}[2]{{%
\expandafter\ifx\csname l@#1\endcsname\relax
\typeout{** WARNING: IEEEtran.bst: No hyphenation pattern has been}%
\typeout{** loaded for the language `#1'. Using the pattern for}%
\typeout{** the default language instead.}%
\else
\language=\csname l@#1\endcsname
\fi
#2}}
\providecommand{\BIBdecl}{\relax}
\BIBdecl

\bibitem{Kadosh_2023}
\BIBentryALTinterwordspacing
T.~Kadosh, N.~Hasabnis, T.~Mattson, Y.~Pinter, and G.~Oren, ``Quantifying openmp: Statistical insights into usage and adoption,'' in \emph{2023 IEEE High Performance Extreme Computing Conference (HPEC)}.\hskip 1em plus 0.5em minus 0.4em\relax IEEE, Sep. 2023, p. 1–7. [Online]. Available: \url{http://dx.doi.org/10.1109/HPEC58863.2023.10363459}
\BIBentrySTDinterwordspacing

\bibitem{intel_inspector}
\BIBentryALTinterwordspacing
Intel, ``Inspector,'' n.d., accessed: 2024-12-18. [Online]. Available: \url{https://www.intel.com/content/www/us/en/developer/tools/oneapi/ inspector.html}
\BIBentrySTDinterwordspacing

\bibitem{serebryany2009threadsanitizer}
K.~Serebryany and T.~Iskhodzhanov, ``Threadsanitizer: data race detection in practice,'' in \emph{Proceedings of the workshop on binary instrumentation and applications}, 2009, pp. 62--71.

\bibitem{Helgrind+}
A.~Jannesari, K.~Bao, V.~Pankratius, and W.~F. Tichy, ``Helgrind+: An efficient dynamic race detector,'' in \emph{2009 IEEE International Symposium on Parallel \& Distributed Processing}, 2009, pp. 1--13.

\bibitem{helgrindplus-lib}
A.~Jannesari and W.~F. Tichy, ``Library-independent data race detection,'' \emph{IEEE Transactions on Parallel and Distributed Systems}, vol.~25, no.~10, pp. 2606--2616, 2014.

\bibitem{Bora/taco/2020}
\BIBentryALTinterwordspacing
U.~Bora, S.~Das, P.~Kukreja, S.~Joshi, R.~Upadrasta, and S.~Rajopadhye, ``{LLOV: A Fast Static Data-Race Checker for OpenMP Programs},'' \emph{ACM Trans. Archit. Code Optim.}, vol.~17, no.~4, Dec. 2020. [Online]. Available: \url{https://doi.org/10.1145/3418597}
\BIBentrySTDinterwordspacing

\bibitem{openrace}
B.~Swain, B.~Liu, P.~Liu, Y.~Li, A.~Crump, R.~Khera, and J.~Huang, ``Openrace: An open source framework for statically detecting data races,'' in \emph{2021 IEEE/ACM 5th International Workshop on Software Correctness for HPC Applications (Correctness)}, 2021, pp. 25--32.

\bibitem{polly}
\BIBentryALTinterwordspacing
T.~GROSSER, A.~GROESSLINGER, and C.~LENGAUER, ``Polly — performing polyhedral optimizations on a low-level intermediate representation,'' \emph{Parallel Processing Letters}, vol.~22, no.~04, p. 1250010, 2012. [Online]. Available: \url{https://doi.org/10.1142/S0129626412500107}
\BIBentrySTDinterwordspacing

\bibitem{LLVM_AA}
\BIBentryALTinterwordspacing
LLVM, ``Alias analysis infrastructure - llvm 20.0.0git documentation,'' 2024. [Online]. Available: \url{https://llvm.org/docs/AliasAnalysis.html}
\BIBentrySTDinterwordspacing

\bibitem{dataracebench}
\BIBentryALTinterwordspacing
C.~Liao, P.-H. Lin, J.~Asplund, M.~Schordan, and I.~Karlin, ``Dataracebench: a benchmark suite for systematic evaluation of data race detection tools,'' in \emph{Proceedings of the International Conference for High Performance Computing, Networking, Storage and Analysis}, ser. SC '17.\hskip 1em plus 0.5em minus 0.4em\relax New York, NY, USA: Association for Computing Machinery, 2017. [Online]. Available: \url{https://doi.org/10.1145/3126908.3126958}
\BIBentrySTDinterwordspacing

\bibitem{pathG}
\BIBentryALTinterwordspacing
F.~Yu, S.-C. Yang, F.~Wang, G.-C. Chen, and C.-C. Chan, ``Symbolic consistency checking of openmp parallel programs,'' \emph{SIGPLAN Not.}, vol.~47, no.~5, p. 139–148, Jun. 2012. [Online]. Available: \url{https://doi.org/10.1145/2345141.2248438}
\BIBentrySTDinterwordspacing

\bibitem{ompdart}
\BIBentryALTinterwordspacing
L.~Marzen, A.~Dutta, and A.~Jannesari, ``Static generation of efficient openmp offload data mappings,'' in \emph{Proceedings of the International Conference for High Performance Computing, Networking, Storage, and Analysis}, ser. SC '24.\hskip 1em plus 0.5em minus 0.4em\relax IEEE Press, 2024. [Online]. Available: \url{https://doi.org/10.1109/SC41406.2024.00041}
\BIBentrySTDinterwordspacing

\bibitem{z3SMT}
L.~De~Moura and N.~Bj\o{}rner, ``Z3: an efficient smt solver,'' in \emph{Proceedings of the Theory and Practice of Software, 14th International Conference on Tools and Algorithms for the Construction and Analysis of Systems}, ser. TACAS'08/ETAPS'08.\hskip 1em plus 0.5em minus 0.4em\relax Berlin, Heidelberg: Springer-Verlag, 2008, p. 337–340.

\bibitem{LeChenDataRaceLLM}
\BIBentryALTinterwordspacing
L.~Chen, X.~Ding, M.~Emani, T.~Vanderbruggen, P.-H. Lin, and C.~Liao, ``Data race detection using large language models,'' in \emph{Proceedings of the SC '23 Workshops of The International Conference on High Performance Computing, Network, Storage, and Analysis}, ser. SC-W '23.\hskip 1em plus 0.5em minus 0.4em\relax New York, NY, USA: Association for Computing Machinery, 2023, p. 215–223. [Online]. Available: \url{https://doi.org/10.1145/3624062.3624088}
\BIBentrySTDinterwordspacing

\bibitem{openai2024gpt4technicalreport}
\BIBentryALTinterwordspacing
OpenAI, ``Gpt-4 technical report,'' 2024. [Online]. Available: \url{https://arxiv.org/abs/2303.08774}
\BIBentrySTDinterwordspacing

\bibitem{li2023starcodersourceyou}
\BIBentryALTinterwordspacing
{R. Li et al.}, ``Starcoder: may the source be with you!'' 2023. [Online]. Available: \url{https://arxiv.org/abs/2305.06161}
\BIBentrySTDinterwordspacing

\bibitem{Llama2}
\BIBentryALTinterwordspacing
{H. Touvron et al.}, ``Llama 2: Open foundation and fine-tuned chat models,'' 2023. [Online]. Available: \url{https://arxiv.org/abs/2307.09288}
\BIBentrySTDinterwordspacing

\bibitem{Hallcination-Haung}
\BIBentryALTinterwordspacing
L.~Huang, W.~Yu, W.~Ma, W.~Zhong, Z.~Feng, H.~Wang, Q.~Chen, W.~Peng, X.~Feng, B.~Qin, and T.~Liu, ``A survey on hallucination in large language models: Principles, taxonomy, challenges, and open questions,'' \emph{ACM Transactions on Information Systems}, Nov. 2024. [Online]. Available: \url{http://dx.doi.org/10.1145/3703155}
\BIBentrySTDinterwordspacing

\bibitem{Xu2024HallucinationII}
\BIBentryALTinterwordspacing
Z.~Xu, S.~Jain, and M.~S. Kankanhalli, ``Hallucination is inevitable: An innate limitation of large language models,'' \emph{ArXiv}, vol. abs/2401.11817, 2024. [Online]. Available: \url{https://api.semanticscholar.org/CorpusID:267069207}
\BIBentrySTDinterwordspacing

\bibitem{Hallucination3}
\BIBentryALTinterwordspacing
N.~M. Guerreiro, E.~Voita, and A.~Martins, ``Looking for a needle in a haystack: A comprehensive study of hallucinations in neural machine translation,'' in \emph{Proceedings of the 17th Conference of the European Chapter of the Association for Computational Linguistics}, A.~Vlachos and I.~Augenstein, Eds.\hskip 1em plus 0.5em minus 0.4em\relax Dubrovnik, Croatia: Association for Computational Linguistics, May 2023, pp. 1059--1075. [Online]. Available: \url{https://aclanthology.org/2023.eacl-main.75}
\BIBentrySTDinterwordspacing

\bibitem{openmp60}
\BIBentryALTinterwordspacing
{OpenMP Architecture Review Board}, ``Openmp application programming interface version 6.0,'' OpenMP Architecture Review Board, 2024, accessed: 2024-12-20. [Online]. Available: \url{https://www.openmp.org/wp-content/uploads/OpenMP-API-Specification-6-0.pdf}
\BIBentrySTDinterwordspacing

\bibitem{Mahmud2023AUTOPARLLMGA}
\BIBentryALTinterwordspacing
Q.~I. Mahmud, A.~TehraniJamsaz, H.~D. Phan, L.~Chen, N.~K. Ahmed, and A.~Jannesari, ``Autoparllm: Gnn-guided automatic code parallelization using large language models,'' \emph{ArXiv}, vol. abs/2310.04047, 2023. [Online]. Available: \url{https://api.semanticscholar.org/CorpusID:263831390}
\BIBentrySTDinterwordspacing

\bibitem{Discopop}
Z.~Li, R.~Atre, Z.~Ul-Huda, A.~Jannesari, and F.~Wolf, ``Discopop: A profiling tool to identify parallelization opportunities,'' in \emph{Tools for High Performance Computing 2014}, C.~Niethammer, J.~Gracia, A.~Kn{\"u}pfer, M.~M. Resch, and W.~E. Nagel, Eds.\hskip 1em plus 0.5em minus 0.4em\relax Cham: Springer International Publishing, 2015, pp. 37--54.

\bibitem{Pattern2}
Z.~U. Huda, R.~Atre, A.~Jannesari, and F.~Wolf, ``Automatic parallel pattern detection in the algorithm structure design space,'' in \emph{2016 IEEE International Parallel and Distributed Processing Symposium (IPDPS)}, 2016, pp. 43--52.

\bibitem{pattern3}
\BIBentryALTinterwordspacing
Z.~Li, R.~Atre, Z.~Huda, A.~Jannesari, and F.~Wolf, ``Unveiling parallelization opportunities in sequential programs,'' \emph{J. Syst. Softw.}, vol. 117, no.~C, p. 282–295, Jul. 2016. [Online]. Available: \url{https://doi.org/10.1016/j.jss.2016.03.045}
\BIBentrySTDinterwordspacing

\bibitem{autmaticConstSelect}
\BIBentryALTinterwordspacing
M.~Norouzi, F.~Wolf, and A.~Jannesari, ``Automatic construct selection and variable classification in openmp,'' in \emph{Proceedings of the ACM International Conference on Supercomputing}, ser. ICS '19.\hskip 1em plus 0.5em minus 0.4em\relax New York, NY, USA: Association for Computing Machinery, 2019, p. 330–341. [Online]. Available: \url{https://doi.org/10.1145/3330345.3330375}
\BIBentrySTDinterwordspacing

\bibitem{Daikon}
\BIBentryALTinterwordspacing
M.~D. Ernst, J.~H. Perkins, P.~J. Guo, S.~McCamant, C.~Pacheco, M.~S. Tschantz, and C.~Xiao, ``The daikon system for dynamic detection of likely invariants,'' \emph{Science of Computer Programming}, vol.~69, no.~1, pp. 35--45, 2007, special issue on Experimental Software and Toolkits. [Online]. Available: \url{https://www.sciencedirect.com/science/article/pii/S016764230700161X}
\BIBentrySTDinterwordspacing

\bibitem{perfograph}
A.~T. Jamsaz, Q.~I. Mahmud, L.~Chen, N.~K. Ahmed, and A.~Jannesari, ``Perfograph: a numerical aware program graph representation for performance optimization and program analysis,'' in \emph{Proceedings of the 37th International Conference on Neural Information Processing Systems}, ser. NIPS '23.\hskip 1em plus 0.5em minus 0.4em\relax Red Hook, NY, USA: Curran Associates Inc., 2024.

\end{thebibliography}

\end{document}